\documentclass[aps,prx,twocolumn,superscriptaddress,longbibliography]{revtex4-2}

\usepackage[colorlinks=true,linkcolor=blue,citecolor=blue,urlcolor=blue]{hyperref} 
\usepackage{amsmath}
\usepackage{amssymb}
\usepackage{physics}
\usepackage{graphicx}
\usepackage{siunitx}
\usepackage{color}

\definecolor{green1}{rgb}{0.1,0.65,0.0}
\definecolor{green2}{rgb}{0.1,0.65,0.1}

\begin{document}
\title{Deterministic Control of Photon-Number Probabilities\\via Phase-Controlled Quantum Interference}

\author{Sang Kyu Kim}
\email{sangkyu.kim@tum.de}
\affiliation{Walter Schottky Institut, TUM School of Computation, Information and Technology, and MCQST, Technische Universit\"at M\"unchen, 85748 Garching, Germany}%
\affiliation{Institute for Advanced Study, Technische Universit\"at M\"unchen, 85748 Garching, Germany}%

\author{Eduardo Zubizarreta Casalengua}
\affiliation{Walter Schottky Institut, TUM School of Computation, Information and Technology, and MCQST, Technische Universit\"at M\"unchen, 85748 Garching, Germany}%

\author{Yeji Sim}
\affiliation{Walter Schottky Institut, TUM School of Computation, Information and Technology, and MCQST, Technische Universit\"at M\"unchen, 85748 Garching, Germany}

\author{\\Friedrich Sbresny}
\affiliation{Walter Schottky Institut, TUM School of Computation, Information and Technology, and MCQST, Technische Universit\"at M\"unchen, 85748 Garching, Germany}%

\author{Carolin Calcagno}
\affiliation{Walter Schottky Institut, TUM School of Computation, Information and Technology, and MCQST, Technische Universit\"at M\"unchen, 85748 Garching, Germany}%

\author{Hubert Riedl}
\affiliation{Walter Schottky Institut, TUM School of Natural Sciences, and MCQST, Technische Universit\"at M\"unchen, 85748 Garching, Germany}%

\author{Jonathan J. Finley}
\affiliation{Walter Schottky Institut, TUM School of Natural Sciences, and MCQST, Technische Universit\"at M\"unchen, 85748 Garching, Germany}

\author{\\Elena {del Valle}}
\affiliation{Departamento de F\'isica Te\'orica de la Materia Condensada, Universidad Aut\'onoma de Madrid, 28049 Madrid, Spain}
\affiliation{Institute for Advanced Study, Technische Universit\"at M\"unchen, 85748 Garching, Germany}
\affiliation{Condensed Matter Physics Center (IFIMAC),Universidad Aut\'onoma de Madrid, 28049 Madrid, Spain}

\author{Carlos~Ant\'on-Solanas}
\affiliation{Departamento de F\'isica de Materiales and Instituto Nicol\'as Cabrera, Universidad Aut\'onoma de Madrid, 28049 Madrid, Spain}
\affiliation{Condensed Matter Physics Center (IFIMAC),Universidad Aut\'onoma de Madrid, 28049 Madrid, Spain}

\author{Kai M\"uller}
\email{kai.mueller@tum.de}
\affiliation{Walter Schottky Institut, TUM School of Computation, Information and Technology, and MCQST, Technische Universit\"at M\"unchen, 85748 Garching, Germany}%

\author{Lukas Hanschke}
\email{lukas.hanschke@tum.de}
\affiliation{Walter Schottky Institut, TUM School of Computation, Information and Technology, and MCQST, Technische Universit\"at M\"unchen, 85748 Garching, Germany}%

\date{\today}
\begin{abstract}
Deterministically tailoring optical Fock states beyond the single-photon level is crucial for boson sampling, loss-tolerant photonic qubits, and quantum-enhanced sensing, however has yet remained elusive.  
Here, we report an all-linear-optical protocol that converts a resonantly driven single-photon emitter into a deterministic generator of vacuum---single-photon---two-photon states.
A phase-stabilized, path-unbalanced Mach-Zehnder interferometer combines vacuum---single-photon interference and Hong-Ou-Mandel effect, providing two knobs to shape photon-number probabilities. 
By tuning these knobs, we observe a dynamic transition from antibunching to strong bunching in correlation measurements.
A fully quantum-mechanical, discrete time-bin model maps these results onto the tailored photon statistics.
The same framework predicts that two indistinguishable emitters would extend the accessible space to deterministic NOON states and single-photon filtering.  
This protocol relying on linear optics and available single-photon sources provides a scalable, chip-compatible, and platform-independent route to on-demand and deterministic few-photon resources for quantum metrology, photonic computing, as well as long-distance quantum networks.

\end{abstract}

\maketitle

\begin{figure*}[t]
  \centering
  \includegraphics[width=\linewidth]{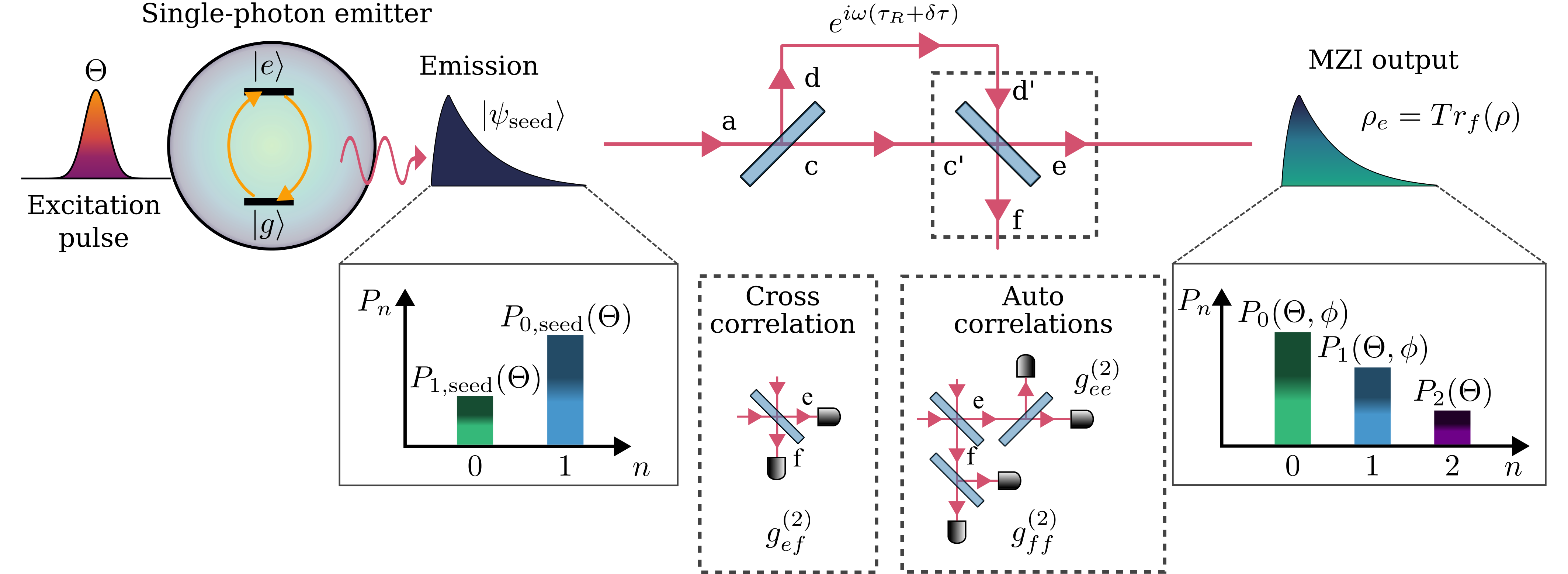}
    \caption{
    Experimental concept for deterministic photon-number probability control. 
    A resonant pulse with pulse area $\Theta$ coherently excites a single-photon emitter, preparing $\ket{\psi_{\mathrm{seed}}}$ 
    whose photon-number probabilities $P_{0,\mathrm{seed}}$ and~$P_{1,\mathrm{seed}}$ are determined by~$\Theta$~(left inset). Consecutively emitted pulses enter a phase-stabilized, path-unbalanced MZI. Interference between adjacent pulses with relative phase $\phi=\omega\delta\tau$ generates a joint state~$\rho$ that is entangled in the time-bin and path degrees of freedom.
    Tracing over output $f$ gives $\rho_e$, whose diagonal elements $P_0$, $P_1$, and $P_2$ are the photon-number probabilities at output $e$.
    While $P_2$ is fixed by $\Theta$, $P_0$ and $P_1$ vary with $\Theta$ and $\phi$ (right inset), enabling deterministic tailoring of photon-number statistics with linear optics alone.
    The second-order cross-correlation $g^{(2)}_{ef}$ and the autocorrelations $g^{(2)}_{ee(ff)}$ at output $e$($f$), measured with the setups sketched in the dashed center insets, reveal the engineered photon statistics.
    }
  \label{fig:exp}
\end{figure*}

\section{Introduction}

Non-classical light sources that deliver on-demand, high-purity single photons are a cornerstone of quantum technologies ranging from quantum networks to photonic quantum computing and quantum-enhanced metrology~\cite{gisin2002quantum, giovannetti2004quantum, Sparrow2018, obrien2009photonic, kok2007linear, slussarenko2019photonic}. Yet many advanced protocols, such as boson sampling~\cite{broome2013photonic, Zhong2020}, loss-tolerant photonic qubits~\cite{varnava2008how, ralph2005efficient}, and sub-shot-noise sensing~\cite{caves1981quantum, pezze2018quantum} require deterministic control over optical Fock-state superpositions that extend beyond the single-photon manifold. Precise control over the vacuum~$\ket{0}$ and single-photon~$\ket{1}$ states has been demonstrated using resonantly driven single quantum emitters, such as semiconductor quantum dots~(QDs) with high purity and indistinguishability~\cite{Senellart2017, Hanschke2018, Schweickert2018, He2013, Schoell2019, Ding2025}. However, accessing higher multiphoton states like the two-photon state $\ket{2}$ typically requires probabilistic nonlinear processes or suffers from intrinsic multiphoton emission noise, which degrades the state fidelity and hampers scalability, limiting the practical realization of scalable quantum photonic circuits~\cite{wang2019boson, uppu2020scalable}.

Recent efforts have explored the coherent generation of two-photon states using resonance fluorescence from a single quantum emitter in the detuned Heitler regime~\cite{masters23a,liu24a}, but the overall two-photon probability remains far below that of the one-photon component and the photons are frequency-distinguishable. A related approach, ``re-excitation'' can yield temporally close photon pairs~\cite{Fischer2017,loredo2019generation}, yet the photons are distinguishable in time and the two-photon probability is fundamentally tied to the emitter decay time and pulse length, leading to a state $\ket{1,1}$ instead of $\ket{2}$. Most recently, Liu \emph{et al.} have demonstrated cavity-enhanced spontaneous two-photon emission from a cascade system in a coherent manner, realizing polarization-entangled photon pairs~\cite{liu2024quantum} but with limited control over the vacuum and single-photon components.

A simple, alternative way is to harness linear-optical interference of indistinguishable photons. Mach-Zehnder interferometers (MZIs) are routinely employed to benchmark photon indistinguishability~\cite{santori2002indistinguishable} and theory predicts that, with appropriate input states, a MZI can function as a deterministic quantum processor for photon-number superpositions~\cite{Campos1989,knill2001scheme,OBrien2003,kiraz2004, Spagnolo2022}. Translating that theoretical promise into a practical, phase-stable, pulsed experiment---and validating it against photon-number probability---remains an open task for the realization of deterministic linear-optical quantum information processing~\cite{Bartolucci2023, rudolph2017optimistic}.

In this work, we demonstrate deterministic control of photon-number probabilities for up to two photons through all-linear optics and non-classical light---a phase-controlled MZI fed by a coherent single-photon source. Experimentally, we combine the MZI with a resonantly driven QD that emits highly indistinguishable single photons. By tuning the excitation pulse area $\Theta$ and the interferometric phase $\phi$, we deterministically tailor the probabilities $P_0$, $P_1$, and $P_2$ for the vacuum, single-, and two-photon components. This is manifested in correlation measurements that show a transition from antibunching to bunching.
Our fully quantum-mechanical description of the path-unbalanced MZI, combined with sequential pulses in a discrete time-bin model, allows us to reconstruct the engineered photon-number probability landscape from photon correlations.
Finally, we confirm that our scheme delivers phase-tunable, two-photon-containing states with negligible higher-order components, all without nonlinear optics or probabilistic heralding sources. Our results provide a scalable pathway for engineering few-photon resources for quantum information, sensing, and networking.


\section{Mach-Zehnder interferometer with quantum light}
Fig.~\ref{fig:exp} shows the proposed protocol: a phase-stabilized MZI that combines  
(i) vacuum---single-photon interference~\cite{loredo2019generation} and  
(ii) two-photon Hong-Ou-Mandel (HOM) interference~\cite{hong1987},  
thereby enabling coherent, deterministic control of photon-number probabilities.
Any single-photon emitter capable of producing a highly coherent superposition of vacuum and one-photon components can be used in the scheme. 
For instance, a quantum two-level system comprising the ground state $\ket{g}$ and the excited state $\ket{e}$ is resonantly driven by a pulsed laser with adjustable pulse area $\Theta$ and pulse repetition period $\tau_{\mathrm{R}}$. 
This coherent excitation prepares the emission as a seed state $\ket{\psi_{\mathrm{seed}}}$, whose vacuum and single-photon probabilities, $P_{0,\mathrm{seed}}(\Theta)$ and $P_{1,\mathrm{seed}}(\Theta)$, are set by $\Theta$ (left inset). 
This quantum seed is injected into the MZI, where two temporally adjacent pulses interfere at the second beam splitter with relative phase $\phi=\omega\delta\tau$, where $\omega$ is the center frequency of the emission. $\delta\tau$ is the small temporal deviation from the perfect overlap between successive pulses, precisely controlled via a piezoelectric actuator.
This interference produces a path- and time-bin entangled output state $\rho$. Tracing over output $f$ yields the reduced state $\rho_e$ at output $e$. Its diagonal elements $\{P_0,P_1,P_2\}$ define the photon-number probabilities which we tailor (right inset). 
Crucially, the pulse area $\Theta$ fixes the two-photon component $P_2$, while the vacuum and single-photon probabilities $P_0$ and $P_1$ are redistributed jointly by $\Theta$ and the interferometer phase $\phi$. 
The second-order cross-correlation $g^{(2)}_{ef}$ and autocorrelations $g^{(2)}_{ee(ff)}$ of the outputs---interpreted through our time-bin model---map out the photon-number probability distribution (dashed insets).

\begin{figure}[t]
  \centering
  \includegraphics[width=\linewidth]{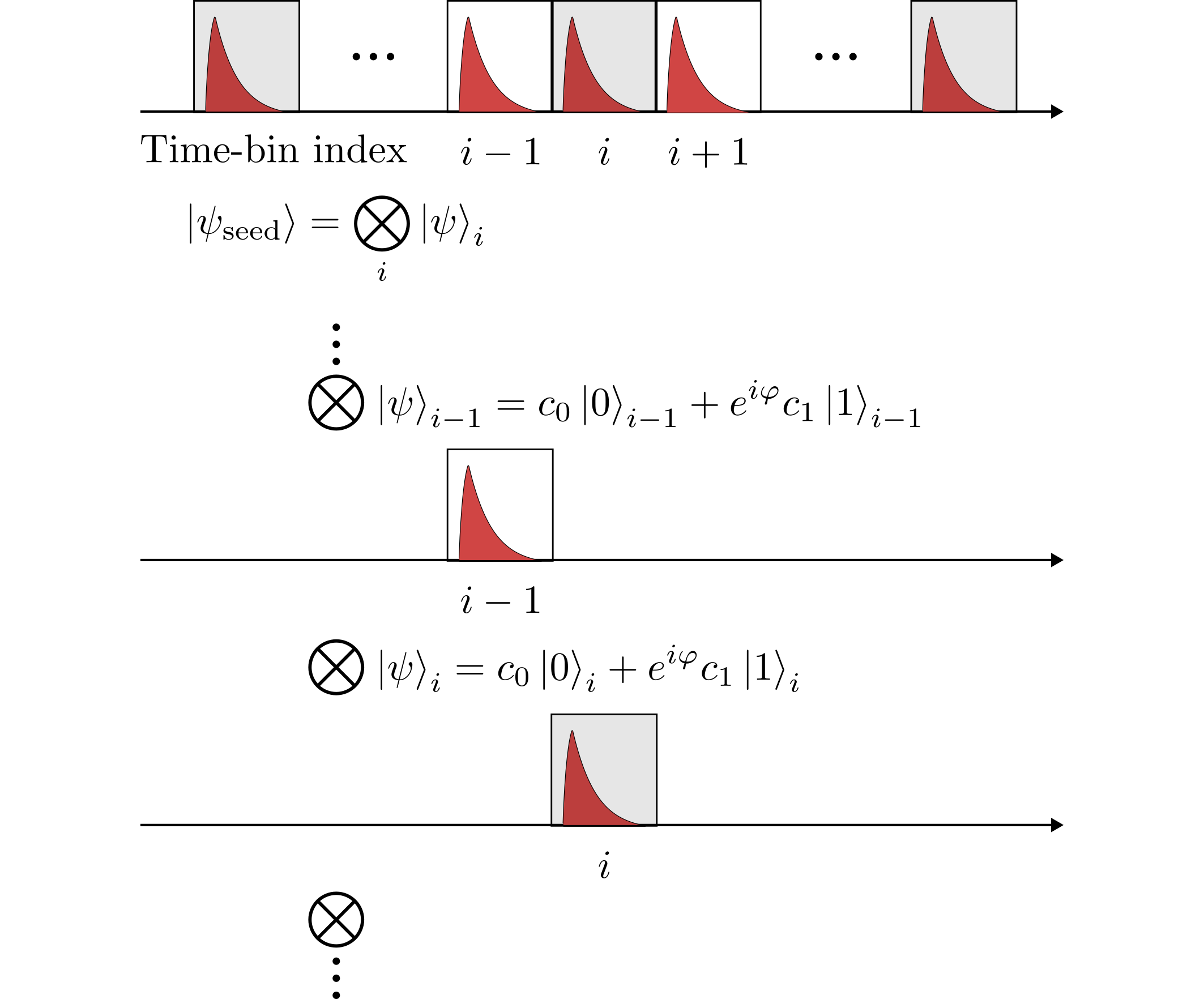}
  \caption{
    Time‑bin representation of the input pulse train. 
    For a periodic pulse train, the time domain is discretized so that each time bin holds at most one emission pulse. 
    The global seed state factorizes as $\ket{\psi_{\mathrm{seed}}}=\otimes_i\ket{\psi}_i$, where $\ket{\psi}_i=c_0\ket0_i+c_1\ket1_i$ describing an arbitrary superposition state of zero- and one-photon Fock states for the identical emission pulse at a time-bin with index $i$.
    }
  \label{fig:timebin}
\end{figure}

\section{Time-Bin Model}
The same unbalanced‑MZI geometry is widely used to characterize the indistinguishability of single-photon sources~\cite{santori2002indistinguishable}. 
In many cases, each emission pulse entering the second beam splitter is considered independently, effectively treating the first beam splitter as a classical, probabilistic component~\cite{loredo2019generation,wells2023coherent}. In fact, the first beam splitter plays a critical role in generating a superposition between the two arms. Combined with the delay in the MZI, it produces path- and time-bin entanglement. Here, we consider both beam splitters fully quantum-mechanically, explicitly accounting for the interaction between successive pulses. 

\subsection{Pulse train in discretized time}
To understand the interference and its impact on the photon-number dynamics, we model a train of pulses in discrete time bins. Each time-bin is assigned to an independent Hilbert space and the full input state is their tensor product.
In our scheme with the pulsed emission of a single photon emitter, the seed injected into the MZI appears as a pulse train illustrated in Fig.~\ref{fig:timebin}. 
When the duration of each emission pulse, which is characterized the radiative lifetime of the excited state for a is short enough compared to the time separation of pulses $\tau_{\mathrm{R}}$, one can define time bins at the pulse repetition period so that each time-bin contains only one pulse without overlap between neighbors.

For simplicity, we describe the seed state in the pure state representation. Our system prepares each emission pulse in an identical state $\ket{\psi_{\mathrm{seed}}}_i = c_0 \ket{0}_i + e^{i\varphi}c_1\ket{1}_i$ at $i$-th time-bin, where $\ket{n}_i$ is n-photon Fock state in the time-bin. The coefficients $c_0$ and $c_1$ are set as $\cos(\Theta/2)$ and $\sin(\Theta/2)$, through Rabi oscillation under pulsed resonant excitation~\cite{allen1975optical}.
The overall optical phase $\varphi$ is irrelevant for our observables and will be omitted below. 
The single emission pulse is assigned to its own Hilbert space, $\ket{\psi_{\mathrm{seed}}}_i \in \mathcal{H}_i$, and the complete seed state belongs to the composite Hilbert space $\ket{\psi_{\mathrm{seed}}} \in \otimes_i\mathcal{H}_i$.
We define pulsed creation operators acting on each spatial mode (shown in Fig.~\ref{fig:exp}) at a time-bin indexed by $i$:
\begin{align}
    \hat{a}_i^\dagger &= \frac{\hat{c}_i^\dagger + \hat{d}_i^\dagger}{\sqrt{2}}, \label{eq:operator_a}\\
    \hat{c}_i^{\prime\dagger} &= \frac{\hat{e}_i^\dagger + \hat{f}_i^\dagger}{\sqrt{2}}, \\
    \hat{d}_i^{\prime\dagger} &= \frac{\hat{e}_i^\dagger - \hat{f}_i^\dagger}{\sqrt{2}}.
\end{align}
Assuming that output $c$ of the first beam splitter is routed to input $c^\prime$ of the second beam splitter without delay, while the longer arm adds a temporal delay of $\tau_R+\delta\tau$, i.e., one time-bin shift $\tau_R$ with an interferometric phase $\phi=\omega\delta\tau$. The operators on mode $c'$ and $d'$ become
\begin{align}
    \hat{c}_i^{\prime\dagger} &= \hat{c}_i^{\dagger}, \\
    \hat{d}_i^{\prime\dagger} &= \hat{d}_{i-1}^{\dagger} e^{i\phi}. \label{eq:operator_d_prime}
\end{align}
Using these operators, we compute the population $n_{e(f),i}$ at output $e$($f$) at time-bin $i$:
\begin{align}
    n_{e(f),i} &= \langle \hat{e}_i^\dagger \hat{e}_i \rangle = \frac{|c_1|^2}{2}\left(1\mp|c_0|^2\cos\phi \right) \label{eq:pop_e}
\end{align}
This expression applies to time bins well inside the pulse train, i.e., where the index $i$ is neither the first nor the last occupied time-bin. In typical measurements, taken in the middle of a long pulse train, the observed counts at the corresponding output are proportional to this quantity.

\subsection{Time-bin correlations}
As we have discretized the time domain and defined the operators, we compute correlations between two modes at time bins $i$ and $j$:
\begin{align}
g^{(2)}_{xy} (\Delta) &= \frac{	\langle \hat{x}_i^\dagger \hat{y}_j^\dagger \hat{y}_j \hat{x}_i \rangle}
{\langle \hat{x}_i^\dagger \hat{x}_i \rangle \langle \hat{y}_j^\dagger \hat{y}_j \rangle},
\label{eq:gen_g2}
\end{align}
with $\Delta=i-j$, where $\hat{x}_i^\dagger(\hat{y}_i^\dagger)$ denotes the creation operator acting on mode $x,y \in \{a, c, c', d, d', e, f\}$ at $i$-th time-bin.
With this, we obtain analytic expressions for second-order autocorrelation and cross-correlation functions. For an ideal system emitting indistinguishable photons described as the pure state $\ket{\psi_{\mathrm{seed}}}_i$, the autocorrelation at output mode $e$ is
\begin{equation}
g^{(2)}_{ee}(\Delta) = 
\begin{cases}
\displaystyle \frac{1}{\left(1 - |c_0|^2 \cos\phi\right)^2}, & \Delta = 0 \\
\displaystyle \frac{3 - 4|c_0|^2 \cos\phi + 2|c_0|^2 \cos(2\phi)}{4\left(1 - |c_0|^2 \cos\phi\right)^2}, & |\Delta| = 1 \label{eq:auto_1}\\
1, & |\Delta| \geq 2,
\end{cases}
\end{equation}
while the cross-correlation reads:
\begin{equation}
g^{(2)}_{ef}(\Delta) = 
\begin{cases}
0, & \Delta = 0 \\[10pt]
\displaystyle \frac{3 -2|c_0|^2 \cos(2\phi)}{4\left(1 - |c_0|^4 \cos^2\phi\right)}, & |\Delta| = 1 \label{eq:cross_1}\\[10pt]
1, & |\Delta| \geq 2.
\end{cases}
\end{equation}
While the correlations above are derived assuming a pure input state, they extend to mixed states by substituting the density matrix of the seed state  $\rho_{\mathrm{seed}, i}$ for time-bin $i$ instead of $\ket{\psi_{\mathrm{seed}}}_i$ into Eq.~\eqref{eq:gen_g2}.
These simplified theoretical time-bin correlations for pure input states reproduce our experimental results once the measured histograms are time-integrated over each pulse period and normalized to the uncorrelated baseline as seen in the following sections.

\section{Experimental Realization}

\begin{figure}[t]
  \centering
  \includegraphics[width=\linewidth]{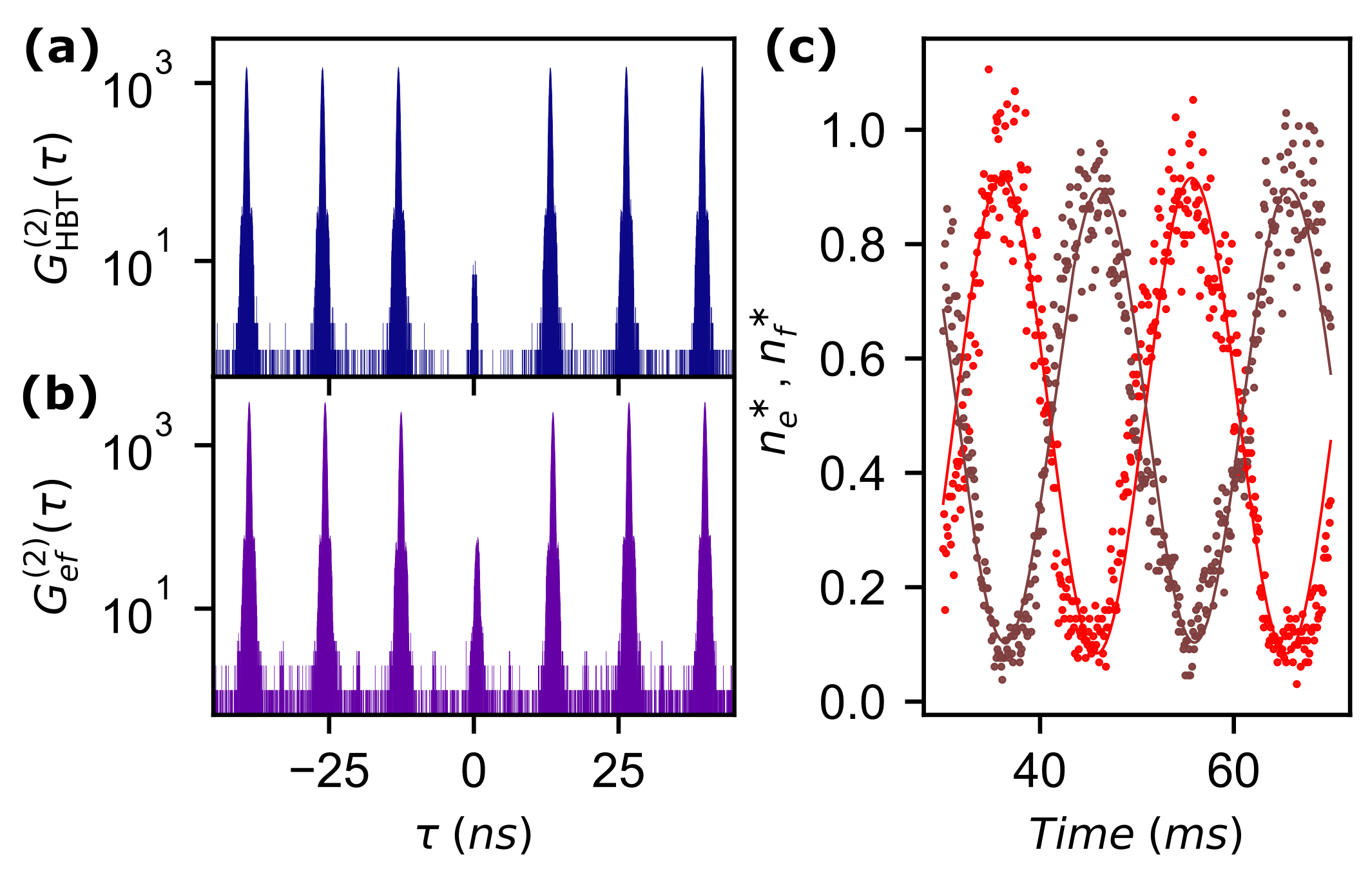}
  \caption{
    Quantum seed characterization.
    (a) Autocorrelation in the HBT setup $G^{(2)}_{\mathrm{HBT}}(\tau)$ of the system emission under resonant $\pi$-pulse excitation.
    The zero-delay peak ($389$ events) is attenuated by a factor of $\sim\!200$ relative to the mean of the side peaks ($\approx\!7.9\times10^{4}$ events), yielding a normalized $g^{(2)}_{\mathrm{HBT}}(0)=0.0049\pm0.0002$.
    (b) Cross-correlation $G^{(2)}_{ef}(\tau)$ measured in the MZI.  
    The central peak ($5205$ events) relative to the uncorrelated average ($\approx\!1.78\times10^{5}$ events) yields an indistinguishability $V_{\mathrm{HOM}}=0.9416\pm0.0006$.
    (c) Normalized counts recorded simultaneously at outputs $e$ (brown) and $f$ (red) of the MIZ while continuously sweeping the interferometer phase $\phi$ in time at pulse area $\Theta = 0.25\pi$. Sinusoidal fits (solid lines) reveal an interference visibility of $0.79$, confirming the high purity of the vacuum---single-photon superposition states seeding the interferometer.
  }
  \label{fig:seed}
\end{figure}

\subsection{Input state preparation: tunable vacuum---single-photon seeds}
We now turn to the experimental findings. We realize a coherently driven single-photon emitter using a neutral exciton in a semiconductor QD excited by resonant laser pulses. For the excitation pulse, we shape femtosecond pulses to have the temporal duration of \SI{10}{\ps} in the intensity full width at half maximum. The pulse repetition period is $\tau_{\mathrm{R}}\!\approx\!\SI{13}{\nano\second}$. This value is considerably longer than the temporal duration of the system emission, thus enabling the application of our time-bin model for this pulsed emission. The experimental details are described in Appendix~\ref{sec:exp}.

In order to deterministically engineer photon-number probabilities, our scheme requires a quantum seed, high-quality single-photon emission with a nearly pure state.
First, we benchmark the single-photon quality of the system emission under resonant $\pi$-pulse excitation. A Hanbury Brown-Twiss (HBT) measurement yields near-perfect antibunching, $g^{(2)}_{\mathrm{HBT}}(0)=0.0049\pm0.0002$, as depicted in Fig.~\ref{fig:seed}(a). The vanishing coincidences at zero-delay in time imply that multiphoton events induced by subsequent excitation within a laser pulse are effectively suppressed for the pulse duration--decay time ratio $\approx\!\SI{10}{\ps}/\SI{220}{\ps}$~\cite{Fischer2017}. 
Routing the same stream through a MZI and recording the cross-correlation, we observe a vanishing central peak in Fig.~\ref{fig:seed}(b). We extract a raw indistinguishability $V_{\mathrm{HOM}}=0.9416\pm0.0006$ which together with the HBT measurement confirms our seed to be near-perfect single-photon emission.

The precise control of photon-number probabilities in our scheme also hinges on seeding the interferometer with coherent and pure vacuum---single-photon superposition states.
We verify the coherence of the seed state by sweeping the interferometer phase and monitoring the single-photon counts at outputs $e$ and $f$.
As described in Eq.~\eqref{eq:pop_e}, the superposition states produce pronounced interference fringes at the population of the outputs, modulated by the vacuum component contribution.
In Fig.~\ref{fig:seed}(c), the normalized counts $n^\ast_e$ and $n^\ast_f$ display interference fringes with visibility  
$v=(n^\ast_e-n^\ast_f)/(n^\ast_e+n^\ast_f)\approx\!0.79$ for a pulse area $\Theta=0.25\pi$. 
This oscillation arises from the phase-dependent vacuum---single-photon interference in the MZI when a highly pure superposition state is considered~\cite{loredo2019generation}.
Together with the high-quality single-photon characteristics, these results confirm a near-unity state purity and strong coherence between temporally adjacent photons, justifying the pure state assumption in our model (see Appendix~\ref{sec:mzi_osc}).  

\begin{figure}
  \centering
  \includegraphics[width=\linewidth]{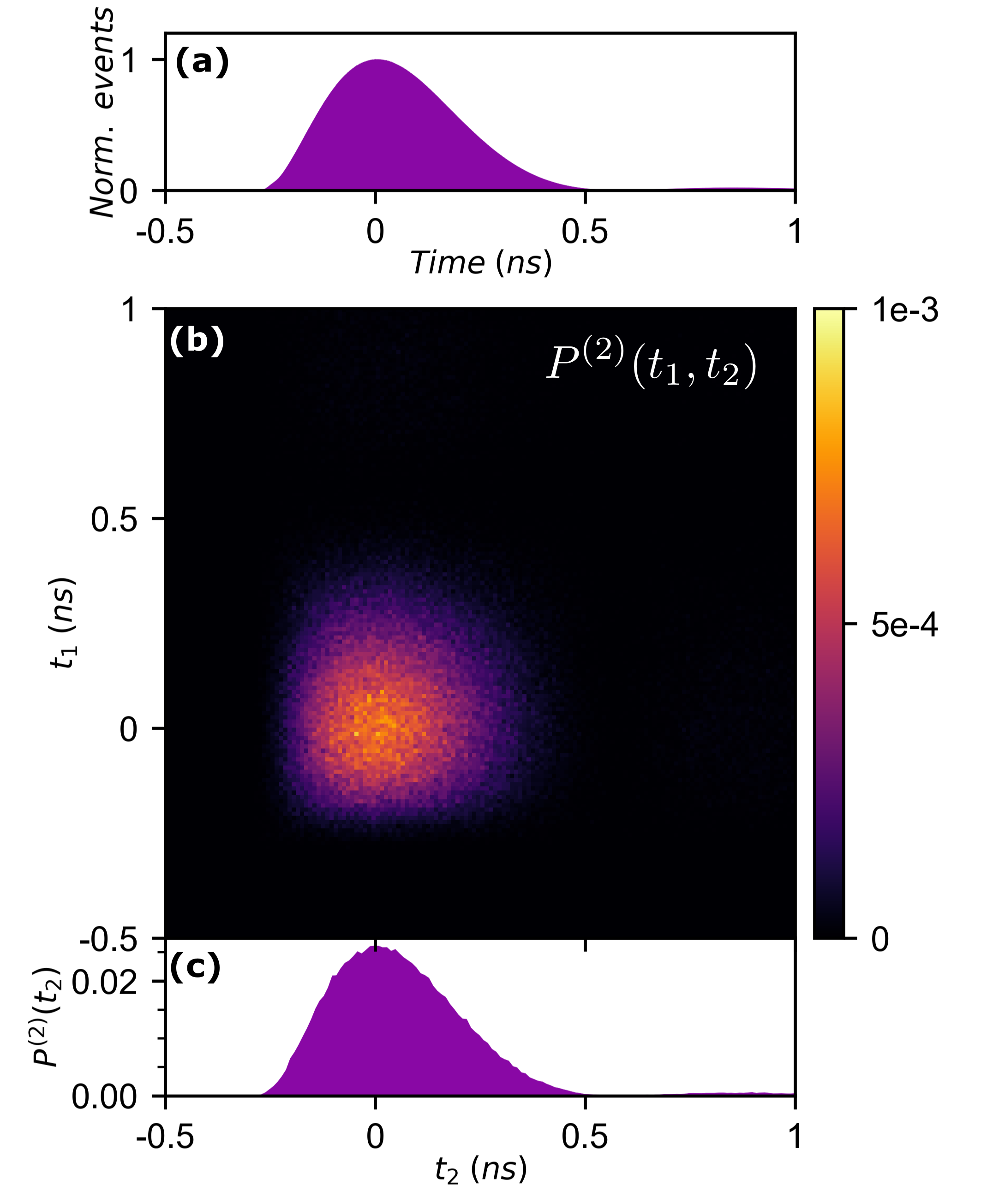}
  \caption{
    Temporal signatures of indistinguishable two-photon pairs at output $e$. 
    (a) Normalized time-resolved photoluminescence, 
    (b) the second-order two-time probability map $P^{(2)}(t_1,t_2)$, and
    (c) its $t_1$-integrated trace of two-photon pairs generated in the MZI. For the seed state, the emitter is driven by $\pi$-pulses, producing a stream of single photons. HOM interference yields two-photon pairs without temporal ordering, visible as a broad central feature spanning the emission wave-packet.
  }
  \label{fig:temporal}
\end{figure}

\subsection{Temporal profile of a genuine two-photon state}
We next investigate the temporal distribution of indistinguishable two-photon pairs arising from HOM interference in the MZI. To achieve the maximum two-photon pair generation, the emitter is driven by resonant $\pi$-pulses. When consecutive pulses simultaneously overlap at the second beam splitter, HOM interference creates a coherent superposition in which both photons exit in the same output but are not bound to any fixed arrival-time ordering. Fig.~\ref{fig:temporal}(a) displays the time-resolved photoluminescence of the output state as single-photon detection events at output $e$ of the MZI are correlated with the synchronization signal of the excitation laser.
The temporal wave-packet exhibits an exponential radiative decay accompanied by an oscillation induced by the fine structure splitting in the cross-polarized detection (see Appendix~\ref{sec:sps}).

To study the two-photon counterpart of the temporal profile, we examine the second-order autocorrelation $G^{(2)}_{ee}(t_1,t_2)$ at output $e$ as a function of the two times $t_1$ and $t_2$, which are the photon arrival times at two single-photon detectors relative to the synchronization signal. In Fig.\ref{fig:temporal}(b), the temporal profile of the two-photon pairs generated from HOM interference is shown as we obtain second-order temporal probability distribution $P^{(2)}(t_1,t_2)$ by normalizing the autocorrelation with the total number of events. 
The two-photon coincidence events form a broad temporal distribution with no tight localization along the two-time axes. After integrating over one time coordinate as depicted in Fig.\ref{fig:temporal}(c), the distribution resembles the temporal wave-packet measured with a single detector in Fig.\ref{fig:temporal}(a), demonstrating that the two-photon temporal probability is spread across the entire emission pulse.
This distribution exhibiting no temporal ordering of the two-photon pairs confirms that the two-photon component arises not from temporally sequential emission, which produces sharp features along the time axes in the two-dimensional map due to a well-defined temporal order of photon generation~\cite{Fischer2017,sbresny2025, jehle2025asymmetrictwophotonresponseincoherently}, but from indistinguishable photons in all degrees of freedom, a genuine $\ket{2}$ state.

\begin{figure*}[t]
  \centering
  \includegraphics[width=\linewidth]{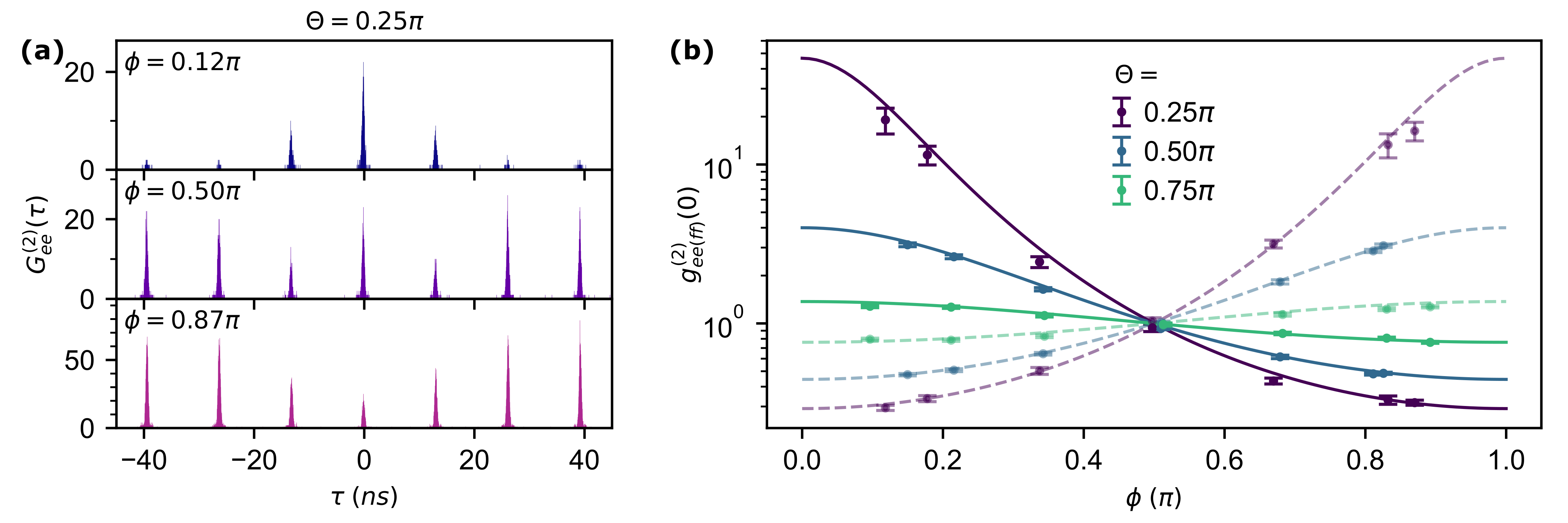}
  \caption{
    Phase-controlled photon statistics.
    (a) Unnormalized second-order autocorrelations $G^{(2)}_{ee}(\tau)$ recorded at a fixed pulse area $\Theta = 0.25\pi$ for three stabilized phases, $\phi = 0.12\pi$, $0.50\pi$, and $0.87\pi$ (from top to bottom panel). The zero-delay peak evolves from strong bunching to antibunching, demonstrating phase-tunable photon statistics.
    (b) Normalized and integrated zero-delay autocorrelations $g^{(2)}_{ee}(0)$ (solid symbols, solid lines) and $g^{(2)}_{ff}(0)$ (semitransparent symbols, dashed lines) versus $\phi$ for three pulse areas $\Theta = 0.25\pi$ (purple), $0.50\pi$ (blue), and $0.75\pi$ (green). Lines and symbols represent theoretical and experimental results, respectively.
  }  
  \label{fig:mzi_g2}
\end{figure*}

\subsection{Deterministic control of photon statistics}
We now study how the two control parameters---the pulse area $\Theta$ and the interferometer phase $\phi$---deterministically reshape the output photon-number statistics.  
Fig.~\ref{fig:mzi_g2}(a) shows unnormalized second-order autocorrelations $G^{(2)}_{ee}(\tau)$ from output $e$ at three stabilized phases while the pulse area is set to $\Theta=0.25\pi$.
The zero-delay peak changes dramatically with $\phi$: strong bunching appears at $\phi = 0.12\pi$ (top panel), whereas a sizable center peak appears at $\phi = 0.50\pi$ (middle panel) and pronounced antibunching is observed at $\phi = 0.87\pi$ (bottom panel).
Integrating each peak and normalizing by the mean of the uncorrelated side peaks yields zero-delay values that swing from $g^{(2)}_{ee}(0)=19.1\pm3.5$ to~$0.31\pm0.01$---a more than $60$-fold change achieved by tuning only the MZI phase $\phi$~\cite{wang2025}.

Fig.~\ref{fig:mzi_g2}(b) extends this analysis across a broader parameter space.  
For three representative pulse areas---$0.25\pi$, $0.50\pi$, and $0.75\pi$---we plot the integrated, normalized autocorrelations at outputs $e$ (solid symbols) and $f$ (semitransparent symbols) as functions of $\phi$.  
The measured results align closely with the theoretical curves at outputs $e$ (solid lines) and $f$ (dashed lines) for an ideal emitter, as shown in the $\Delta =0$ case of Eq.~\eqref{eq:auto_1}. This indicates that the near-optimal quantum seed state preparation from the studied single-photon source and the highly stable optical phase in the interferometer. The observed discrepancies between the theoretical results and the experimental observations can be attributed to several factors, including non-unity indistinguishability and state purity as well as measurement uncertainties such as residual phase fluctuations.
The correlations are $2\pi$-periodic and mirror-symmetric, reflecting the complementary behavior of the two outputs.  
For the pulse area $\Theta$ approaches $\pi$, the phase dependence collapses: both $g^{(2)}_{ee}(0)$ and $g^{(2)}_{ff}(0)$ lock near unity, because the seed state becomes a pure single-photon emission ($P_{0,\mathrm{seed}}\approx\!0$), suppressing phase-dependent vacuum---single-photon interference (see Appendix~\ref{sec:mzi_pi}).

A careful interpretation is required when relating correlation measurements to underlying photon-number probabilities.
Non-monotonic behavior of correlation functions can lead to the common misconception that a higher $g^{(2)}(0)$ directly implies a higher multiphoton probability. 
The strong bunching observed at low $\Theta$ arises from an inflated vacuum component despite small two-photon contribution~\cite{zubizarretacasalengua17a, Grunwald2019}. In fact, the two-photon probability is maximized at $\Theta=\pi$. 
The bunching-to-antibunching transition at a fixed $\Theta$ is governed entirely by phase-controlled vacuum and single-photon components; the two-photon probability $P_2$ remains invariant with phase and is set solely by $\Theta$. 
This insight motivates a comprehensive photon-number probability analysis developed in the next section.

\begin{figure*}[t]
  \centering
  \includegraphics[width=\linewidth]{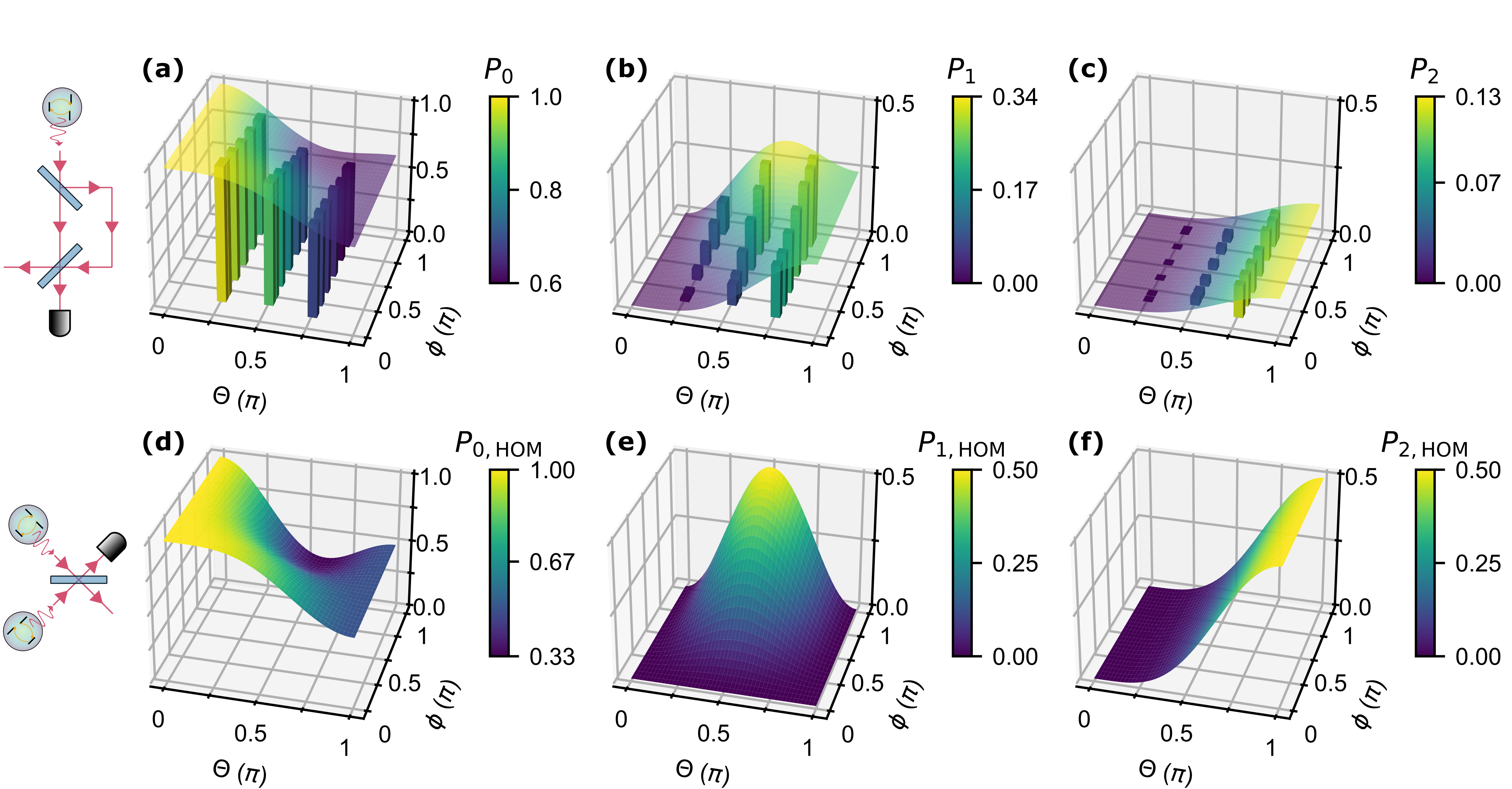}
  \caption{
    Photon-number probability curves in comparison of single-source MZI- and dual-source HOM-type schemes. 
    Theoretically calculated photon-number probabilities as functions of pulse area $\Theta$ and phase $\phi$ are shown as colored surfaces. For the realized single-source setting, the probabilities corresponding to the experimental data in Fig~\ref{fig:mzi_g2} (c) are presented as bar plots.
    (a)--(c) Single-source MZI~(inset): vacuum, single-, and two-photon probabilities $P_0$, $P_1$, and $P_2$ span $P_0\in[0.6,1]$, $P_1\in[0,1/3]$, $P_2\in[0,0.125]$.
    (d)--(f) Dual-source HOM scheme (inset). Output probabilities $P_{\mathrm{HOM},0}$, $P_{\mathrm{HOM},1}$, and $P_{\mathrm{HOM},2}$ expand the accessible space to $P_{\mathrm{HOM},0} \in [1/3, 1]$, $P_{\mathrm{HOM},1} \in [0, 0.5]$, and $P_{\mathrm{HOM},2} \in [0, 0.5]$.
    }
  \label{fig:prob}
\end{figure*} 

\section{Photon-Number Probability Landscape}

\subsection{Single-source MZI}
To quantitatively understand and validate our experimental observations, we compute photon-number probabilities at output $e$ from autocorrelations for pure seed states as~\cite{zubizarretacasalengua17a}
\begin{align}
P_0(\Theta,\phi) =& 1 - P_1(\Theta,\phi) - P_2(\Theta) \label{eq:p0}\\
P_1(\Theta,\phi) =& \frac{1}{4}\sin^4(\Theta/2) \nonumber \\ &+\frac{1}{2} \sin^2(\Theta/2) \cos^2(\Theta/2) (1-\cos\phi) \label{eq:p1}\\
P_2(\Theta) =& \frac{1}{8}\sin^4(\Theta/2). \label{eq:p2}
\end{align}
Since the two outputs $e$ and $f$ differ only by a $\pi$ phase shift, we discuss the results at one output mode without loss of generality. We also restrict the photon-number space to the vacuum, one-, and two-photon manifolds as three-photon events are negligible (see Appendix~\ref{sec:mzi_pi}).
 
Leveraging our validated theoretical model, we systematically explore the accessible photon-number probability landscape enabled by our MZI-based configuration and compare it with a standard HOM interference scheme.
For the experimentally realized case of a MZI fed by a single-photon source, we present the theoretical (surfaces) and experimental photon-number probabilities (bars) in Fig.~\ref{fig:prob}(a)--(c). We extract the probabilities for up to two photons from the autocorrelation measurements, assuming unity collection and detection of the system emission (see Appendix~\ref{sec:extract_prob}). The estimated probabilities align well with the theoretical surfaces across the two-dimensional parameter space, validating deterministic control of photon-number probabilities, albeit with limited accessibility. 

As seen in the curvatures of the probability surfaces, the pulse area $\Theta$ and the phase $\phi$ together shape $P_0$ and $P_1$ in Fig.~\ref{fig:prob}(a) and (b), while $P_2$ in Fig.~\ref{fig:prob}(c) increases monotonically without phase dependency as the pulse area approaches $\pi$. 
The interferometric phase $\phi$ serves as a control knob for tuning the relative weights of the vacuum $P_0$ and single-photon $P_1$ components. This separation of control variables offers practical benefits: one can first set the pulse area to fix the two-photon probability and subsequently adjust the phase to engineer the desired vacuum---single-photon ratio.
We find the accessible ranges, $P_0 \in [0.6, 1]$, $P_1 \in [0, 1/3]$, and $P_2 \in [0, 0.125]$. This limitation results from the fact that the two-photon interference in this configuration does not occur every time-bin due to the nature of a single source. Specifically, when a two-photon state is measured in a time-bin, the composite output state immediately collapses into reduced cases in which two-photon interference cannot occur in the adjacent time bins.

\subsection{Dual-source HOM}
In contrast, HOM interference with two indistinguishable single-photon sources (inset of Fig.~\ref{fig:prob}(d))---or, equivalently, a single source with fast optical demultiplexing~\cite{Lenzini2017, Pont2022, Muenzberg2022}---surpasses the limitation and significantly expands the space of accessible photon-number probability. The dual-source configuration with a single beam splitter eliminates the interaction between successive pulses and thus simplifies the photon-number probabilities. In this setting, they become
\begin{align}
P_{0, \mathrm{HOM}}(\Theta,\phi) =& 1 - P_{1, \mathrm{HOM}}(\Theta,\phi) - P_{2, \mathrm{HOM}}(\Theta)\\
P_{1, \mathrm{HOM}}(\Theta,\phi) =& \sin^2(\Theta/2) \cos^2(\Theta/2) (1-\cos\phi) \\
P_{2, \mathrm{HOM}}(\Theta) =& \frac{1}{2}\sin^4(\Theta/2). \label{eq:p2HOM}
\end{align}
In this case, the achievable probabilities span $P_{0, \mathrm{HOM}} \in [1/3, 1]$, $P_{1, \mathrm{HOM}} \in [0, 0.5]$, and $P_{2, \mathrm{HOM}} \in [0, 0.5]$, as shown in more pronounced curvatures in Fig.~\ref{fig:prob}(d)--(f). Unlike the previous single-source scheme, the two-photon pairs resulting from the two sources do not affect the neighboring time bins. Notably, $P_{1, \mathrm{HOM}}$ can reach $0$ regardless of the pulse area $\Theta$ by controlling the interferometric phase $\phi$. This method with our precise phase control facilitates the purification of the two-photon output state. In particular, it is possible to engineer an output such that it contains exclusively vacuum and two-photon components, independent of the pulse area, by redirecting all single photons to the other output. This is the opposite operation of conventional single-photon filtering, where only single photons appear at an output leveraging cavity-enhanced light-matter interaction~\cite{bennett2016singlephotonfilter, desantis2017singlephotonfilter, tomm24a}. These findings emphasize that our approach, when extended to two single-photon sources with high mutual-indistinguishability~\cite{Beugnon2006, Zhai2022}, unlocks richer non-classical operations such as single-photon filtering and enables deterministic preparation of advanced quantum states such as NOON states.

\section{Conclusion}

We have shown that a single resonantly driven single-photon emitter combined with a phase-controlled, path-unbalanced Mach-Zehnder interferometer enables deterministic and on-demand tailoring of optical Fock-state populations. 
By tuning only two experimental knobs, the pulse area $\Theta$ and the interferometer phase $\phi$, we access photon-number probabilities within the ranges $P_0 \in[0.6,1]$, $P_1\in[0,1/3]$, and $P_2\in[0,0.125]$: $\Theta$ determines the two-photon probability $P_2$, while $\phi$ redistributes the vacuum and single-photon components between the two outputs of the interferometer. 
This tunability manifests itself in a transition of correlations from antibunching to bunching. Our fully quantum-mechanical time-bin model maps these correlations onto the underlying probability landscape, in excellent agreement with experiment. 
Practically, the scheme converts a standard single-photon source into a deterministic generator of vacuum---single-photon---two-photon states, requiring neither post-selection nor heralding.

Looking ahead, extending the protocol to two mutually indistinguishable emitters with a phase-controlled interferometer would expand the reachable space to $P_{2, \mathrm{HOM}}=0.5$, generating deterministic NOON states~\cite{muller2017, kok2002Creation, Walther2004} and to $P_{1, \mathrm{HOM}}=0$ for all $\Theta$, facilitating single-photon filtering on an all-linear-optical platform. This finding highlights a substantial improvement in the control of photon-number probability that can be achieved with multiple indistinguishable emitters that are currently available~\cite{Beugnon2006, Zhai2022}.
Because the entire protocol relies only on coherent state preparation and linear optics, it is platform-independent and readily integrable with chip-scale photonic circuits~\cite{Politi2008,Carolan2015,Harris2017,Wang2020}. We anticipate that the ability to deterministically tailor few-photon states will accelerate boson-sampling experiments, loss-tolerant photonic qubits, and sub-shot-noise metrology, while providing a resource required for scalable, deterministic linear-optical quantum processors, and long-distance quantum networks.

\begin{acknowledgments}
We gratefully acknowledge financial support from the Deutsche Forschungsgemeinschaft (DFG, German Research Foundation) via projects MU 4215/4-1 (CNLG), INST 95/1220-1 (MQCL) and INST 95/1654-1 (PQET), Germany's Excellence Strategy (MCQST, EXC-2111, 390814868). C.A.-S. acknowledges the support from the Comunidad de Madrid fund “Atracci\'on de Talento, Mod. 1”, Ref. 2020-T1/IND-19785, the projects from the Ministerio de Ciencia e Innovaci\'on PID2023-148061NB-I00 and PCI2024-153425, the project ULTRABRIGHT from the Fundaci\'on Ram\'on Areces, the Grant “Leonardo for researchers in Physics 2023” from Fundaci\'on BBVA, and the Spanish State through the Recovery, Transformation, and Resilience Plan (MAD2D-CM-UAM7), and the European Union through the Next Generation EU funds. E.d.V. acknowledges support from the CAM Pricit Plan (Ayudas de Excelencia del Profesorado Universitario), the Technical University of Munich - Institute for Advanced Study (Hans Fischer Fellowship) and the Spanish Ministry of Science, Innovation and Universities through the ``Maria de Maetzu'' Programme for Units of Excellence in R\&D (CEX2023-001316-M), the MCIN/AEI/10.13039/501100011033, FEDER UE, projects No.~PID2020-113415RB-C22 (2DEnLight) and No.~PID2023-150420NB-C31 (Q), and from the Proyecto Sinérgico CAM 2020 Y2020/TCS-6545 (NanoQuCo-CM). 
\end{acknowledgments}

\appendix
\begin{figure*}[t]
  \centering
  \includegraphics[width=\linewidth]{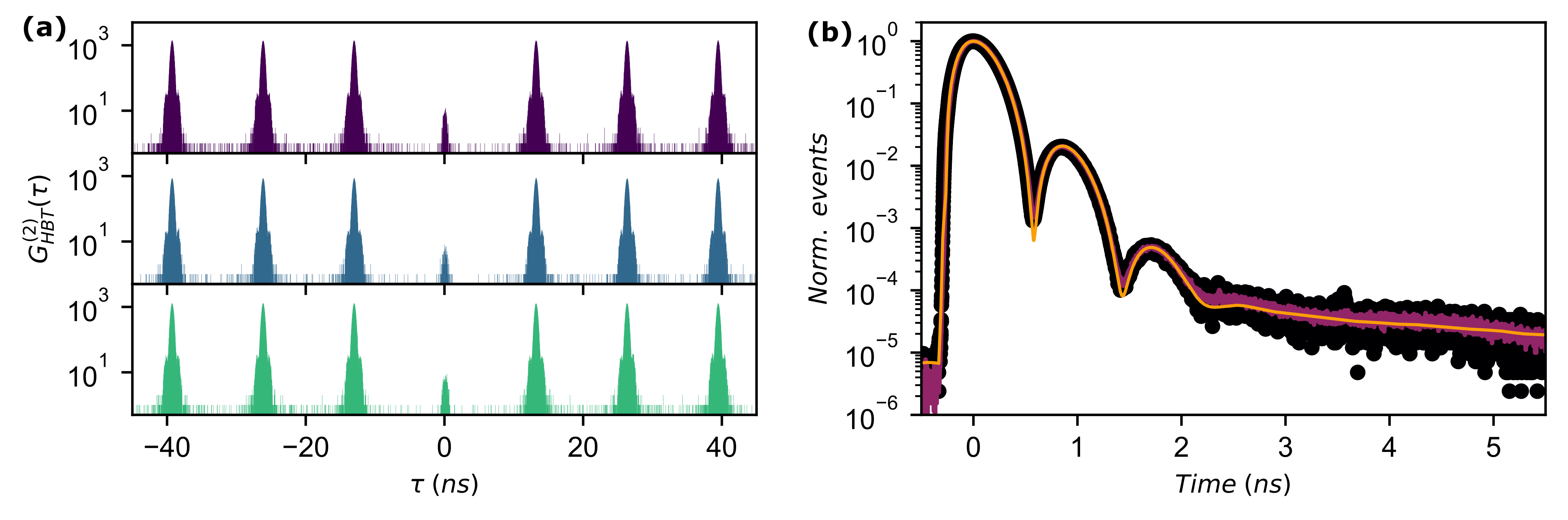}
  \caption{
    Pulsed single-photon characteristics. 
    (a) Autocorrelations in HBT setup $G^{(2)}_{\mathrm{HBT}}(\tau)$ for three different pulse areas $\Theta = 0.25\pi$, $0.50\pi$, and $0.75\pi$ from top to bottom panel.
    (b) Time-resolved photoluminescence of the system emission under a pulse area $\Theta =1\pi$ (black symbols) and its fitting curve (yellow line) exhibit two exponential decays modulated by the fine structure precession. Purple symbols represent the temporal wave-packet after MZI reproduced from Fig.~\ref{fig:temporal}(a) in the main manuscript displaying a perfect overlap.
    }
  \label{fig:QD}
\end{figure*}

\section{Experimental setup}
\label{sec:exp}
The experiments are performed on a single self-assembled InGaAs QD. The sample is cooled to \SI{4.2}{\K} using a helium dip stick. The QD is embedded in a Schottky diode structure, which allows us to stabilize the electronic environment and fine-tune the emission wavelength via the quantum-confined Stark effect~\cite{warburton2000}. Additionally, the structure is grown on top of a distributed Bragg reflector to enhance collection efficiency.
To selectively collect the QD emission and suppress scattered laser light, we employ a cross-polarized resonance fluorescence scheme, where the excitation and detection polarizations are orthogonal~\cite{vamivakas2009}. Further spectral filtering is provided by a transmission grating (\SI{1500}{grooves/mm}) with a \SI{47.6}{\GHz} transmission bandwidth, designed similarly to that in Ref.~\cite{sbresny2025}. In the path-unbalanced MZI in Fig.~\ref{fig:exp}, we control and stabilize the interferometric phase $\phi$ through a piezo-controlled linear stage in the delay arm of the interferometer. Phase stabilization is achieved using software PID control. Specifically, the piezo voltage is updated every \SI{20}{\ms} to maintain the target visibility. For photon-counting and correlation measurements, we use superconducting nanowire single-photon detectors combined with a time-tagging unit.

\section{Characterization of single-photon emitter}
\label{sec:sps}
The single-photon source is realized by resonantly exciting a neutral exciton in the QD with a pulsed Ti:Sa laser, tuned to the emission wavelength of \SI{910}{\nm}. The laser pulses are shaped to have an intensity full width half maximum of \SI{10}{\ps}. The resonant pulsed laser drives the system from the crystal ground state to a superposition of two excited states that are energetically non-degenerate by a fine structure splitting of approximately \SI{600}{\MHz}. When the spectral width of the resonant excitation pulse is much larger than the fine structure splitting, the system can be regarded as an effective two-level system~\cite{Boyle2008}. Futhermore, in the time-bin picture, the fast temporal dynamics, governed by the excited-state superposition, does not play a role once the signal is integrated over the time-bin. The influence of this superposition, however, is visible in time-resolved measurements within a single time bin.

To verify single-photon emission across all pulse areas used, we perform HBT autocorrelation measurements at $\Theta = 0.25\pi$, $0.50\pi$, $0.75\pi$, and $\pi$. 
The results at $\Theta = \pi$ is shown in the top panel of Fig.~\ref{fig:seed}(a) of the main manuscript. In Fig.~\ref{fig:QD}(a), the correlations $G^{(2)}_{HBT}(\tau)$ for $\Theta = 0.25\pi$ (top), $0.50\pi$ (middle), and $0.75\pi$ (bottom) are shown. From these results, we extract $g^{(2)}_{\mathrm{HBT}}(0) = 0.0057 \pm 0.0003$, $0.0052 \pm 0.0003$, and $0.0054 \pm 0.0003$, respectively. These values confirm that multiphoton emission is effectively suppressed for all studied excitation pulse areas.

For $\Theta = \pi$, we further characterize the temporal wave-packet of the single-photon emission. The time-resolved photoluminescence of the system is depicted as black symbols in Fig.~\ref{fig:QD}(b). We fit the data using a model comprising exponential decays and fine structure induced oscillations~\cite{Schoell2019}, convolved with the instrument response function. From the fit, we obtain a major fast radiative decay time of approximately \SI{220}{\ps} and a weaker, long-lived component exceeding \SI{2.5}{\ns}, which we attribute to spin-flip transitions and the resulting dark-exciton recombination~\cite{dalgarno2005dark}.
As a comparison, we also show the temporal wave-packet at an output of the MZI (purple symbols), reproduced from Fig.~\ref{fig:temporal}(a). The excellent overlap between the two profiles evidences that the temporal distribution of delocalized two-photon pairs in Fig.~\ref{fig:temporal}(b) accurately reflects the wave-packet shape of the seed pulses injected into the interferometer.


\begin{figure*}
  \centering
  \includegraphics[width=\linewidth]{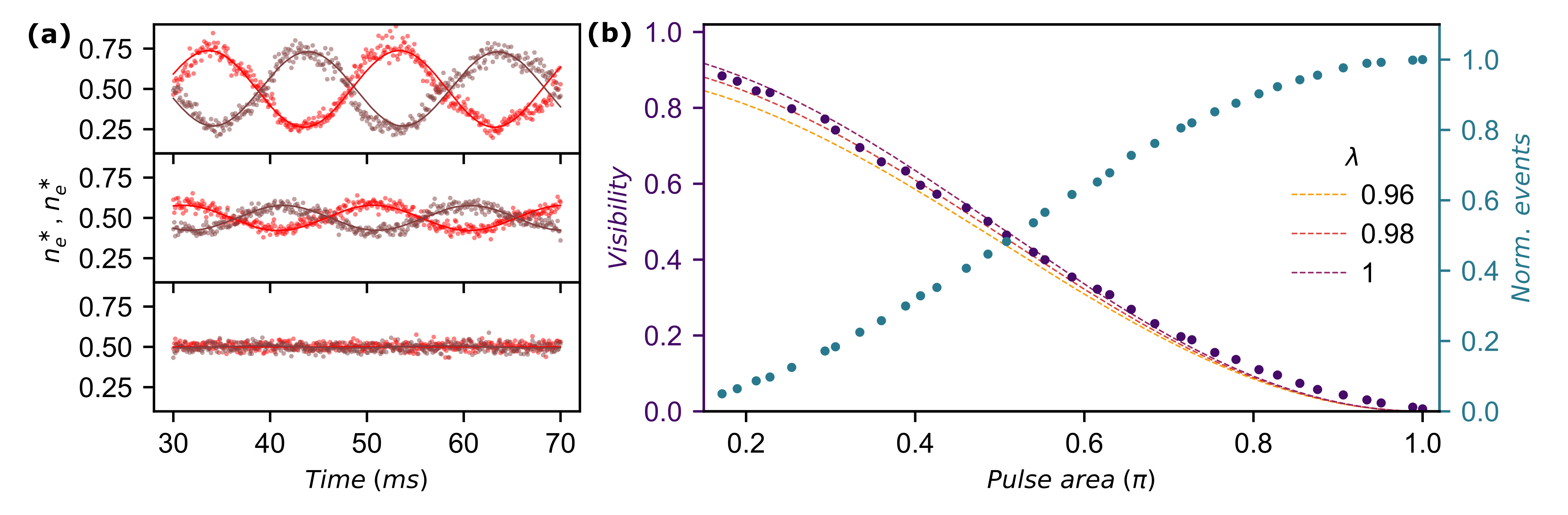}
  \caption{Vacuum---one-photon interference in the MZI.
  (a) Normalized single‑photon counts at outputs $e$ (brown) and $f$ (red) while $\phi$ is scanned, for three pulse areas $\Theta=0.50\pi$, $0.75\pi$, and $\pi$ from top to bottom panel.
  (b) Extracted fringe visibilities (purple) and normalized total detection events (blue) as functions of pulse area.
  Dashed curves show the expected visibility for state purities $\lambda = 0.96$, $0.98$, and $1$.}
  \label{fig:MZI_OSC}
\end{figure*}

\section{Vacuum and one-photon interference in MZI}
\label{sec:mzi_osc}
A clear phase‑dependent interference pattern observed at the two outputs of the MZI verifies that the input pulse train is prepared in a coherent vacuum---one‑photon superposition with high state purity.
The expected population of each output port is given by Eq.~\eqref{eq:pop_e} of the main text.
For a fixed excitation pulse area $\Theta$, we continuously sweep the interferometer phase $\phi$ and record time‑tagged histograms of the detection events at outputs $e$ and $f$.
Fig.~\ref{fig:MZI_OSC}(a) shows the resulting phase‑dependent count traces for three representative pulse areas, $\Theta = 0.50\pi$, $0.75\pi$, and $\pi$ (top to bottom).  
Symbols denote the normalized counts, $n^\ast_{e(f)}(t) = N^\ast_{e(f)}(t) /(\bar{N^\ast_{e}} + \bar{N^\ast_{f}})$, where $N^\ast_{e(f)}(t)$ is the accumulated histogram counts at time $t$, while $\bar{N^\ast_{e}}$ and $\bar{N^\ast_{f}}$ are the corresponding counts averaged over the entire \SI{30}{\second} acquisition window.
Solid lines are fitting curves derived from Eq.~\eqref{eq:pop_e}.
From the fitting, we extract the fringe visibility $v$ as a function of pulse area and plot it as purple symbols in Fig.~\ref{fig:MZI_OSC}(b) together with the normalized total number of detected events (blue).
As expected, the visibility increases as the excitation approaches vanishing driving $\Theta\to0$, where the vacuum component dominates.
Quantitatively, the fringe contrast follows $v = \lambda^2 \sqrt{V_{\mathrm{HOM}}}\,\cos^2(\Theta/2)$, with state purity factor $\lambda$ and indistinguishability $V_{\mathrm{HOM}}$~\cite{loredo2019generation}. 
The extracted visibilities for lower pulse areas constrain the purity to the interval $0.96 \leq \lambda \leq 1$.
As the driving approaches a $\pi$‑pulse, a small deviation emerges, which we attribute to damped Rabi oscillations that prevent unity population inversion, as well as to residual laser scattering and imperfections from the fitting procedure.

\begin{figure}
  \centering
  \includegraphics[width=\linewidth]{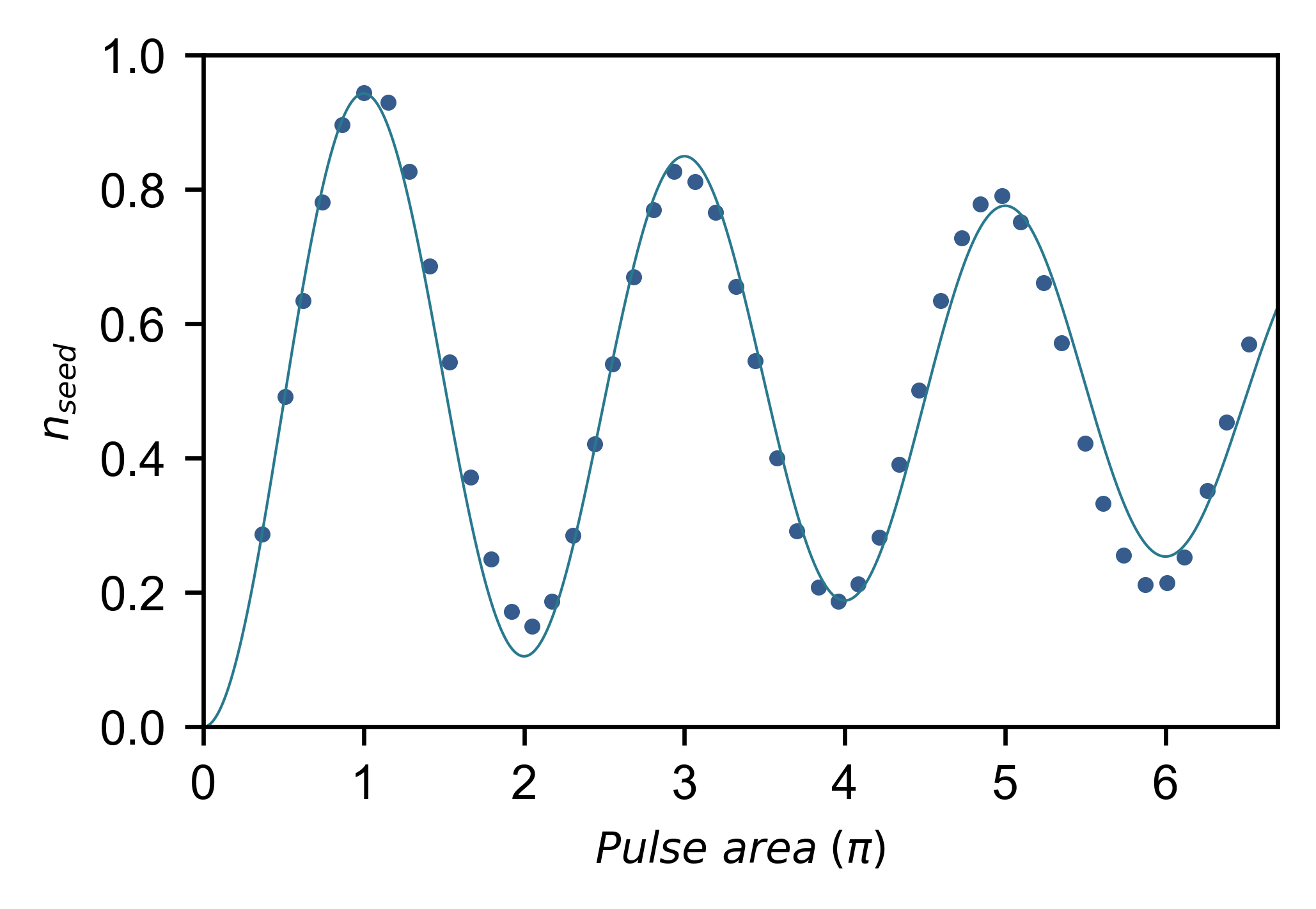}
  \caption{
  Pulse-area-dependent mean photon numbers of the seed state. Assuming a fixed total efficiency, the mean photon numbers (symbols) are evaluated from the normalized emission counts. Under resonant pulsed excitation, coherent population transfer is evidenced by Rabi oscillations of the mean photon numbers as a function of pulse area. Fitting with a spontaneous-emission-induced damping model (line) yields a maximum mean photon number of 0.94 per pulse at $\Theta=\pi$.
  }
  \label{fig:Rabi}
\end{figure}

\begin{figure*}
  \centering
  \includegraphics[width=\linewidth]{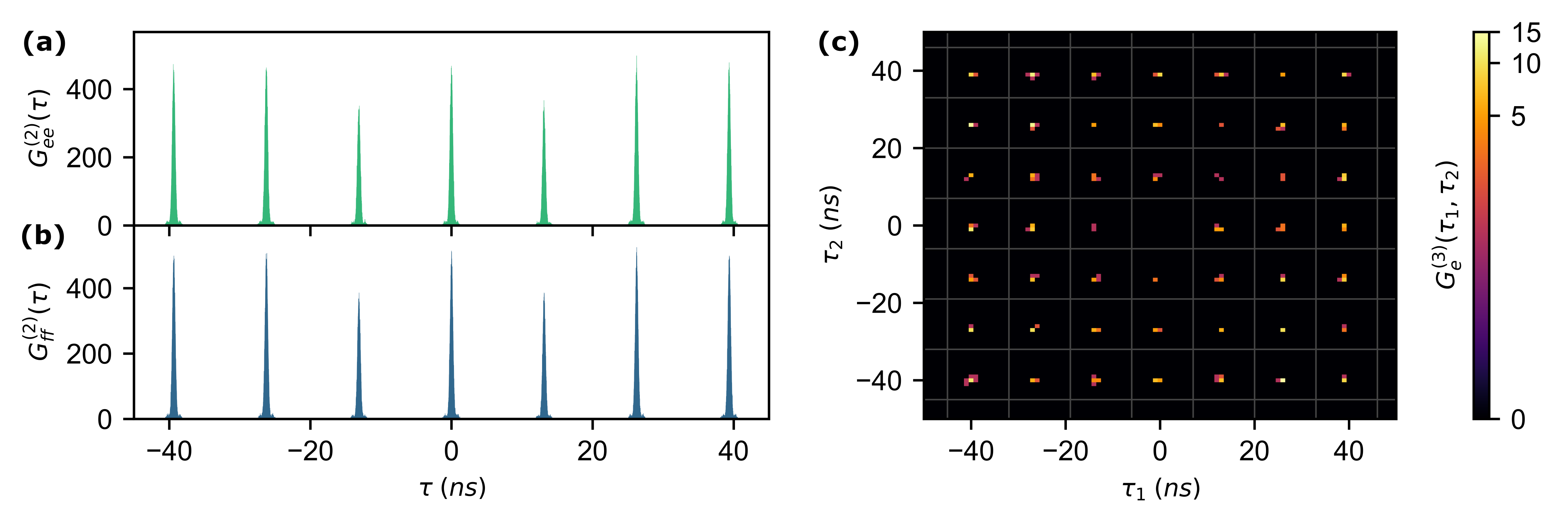}
  \caption{Autocorrelations at the MZI outputs for $\Theta=\pi$. 
  (a) Unnormalized $G^{(2)}{ee}(\tau)$ and (b) $G^{(2)}{ff}(\tau)$, whose zero-delay peaks yield near-unity coincidences after time-bin integration and normalization. In contrast to smaller pulse areas, no phase-dependent behavior is observed.
  (c) Unnormalized third-order correlation $G^{(3)}_{e}(\tau_1,\tau_2)$ at output $e$ for $\Theta=\pi$. The integration boundaries used for the time-bin analysis are indicated by gray lines and are aligned with the pulse period $\tau_R$. In the central bin, $\SI{-6.5}{\ns}\lesssim\tau_{1,2}\lesssim\SI{6.5}{\ns}$, no three-photon coincidences are observed over about $30$ hours, whereas uncorrelated bins show on average $\approx\!11$ events.
  } 
  \label{fig:MZI_pi_g3_g3}
\end{figure*}

\section{Extracting photon-number probabilities}
\label{sec:extract_prob}
The photon-number probabilities of the studied light are extracted, for up to two photons, from second-order autocorrelation measurements, assuming vanishing three-photon events and a known mean photon number~\cite{zubizarretacasalengua17a}. With the HBT measurements of the seed state in Fig.~\ref{fig:QD} (a) and the third-order correlation of the tailored state at output $e$ in Fig.~\ref{fig:MZI_pi_g3_g3} (c), we confirm a negligible three-photon probability for the state of interest at the MZI outputs. To estimate the mean photon number $n_{seed}$ of the seed state, we measure the emission counts as a function of the excitation power, as shown in Fig.~\ref{fig:Rabi}. The pulse area $\Theta$ is proportional to the square root of the excitation power, and we normalize the power so that the value yielding the maximum counts corresponds to $\Theta = \pi$. With a fixed total efficiency, the detected counts are proportional to the mean photon number. We fit the normalized mean photon numbers (symbols), with a theoretical Rabi oscillation curve (line) incorporating spontaneous-emission-induced damping~\cite{loudon2000quantum} as the measured counts exhibit damped Rabi oscillations. 
As the pulse area increases, the oscillation period in the measured data becomes shorter compared to that of the fit, which remains constant. This Rabi-frequency renormalization, induced by longitudinal acoustic phonons in the semiconductor host material~\cite{Ramsay2010renormal}, is not accounted for in our simple model. However, the model agrees reasonably well over the pulse-area range of interest, $\Theta\in[0,\pi]$. Accordingly, we extracted the maximum population inversion of approximately $0.94$ at $\pi$-pulse area.

Assuming negligible non-radiative decay and unit efficiency, about $0.47$ photons per pulse are expected at output $e$. In our experiment, we obtain about $3.18\times10^{-4}$ photons per pulse at $\Theta=\pi$, implying a total efficiency at that output of $\eta_e\!\approx\!6.75\times10^{-4}$. Given this value, we estimate the mean photon number $n_e$ at output $e$ from the measured counts per second $N_e$ and infer the photon-number probabilities from second-order autocorrelation as~\cite{zubizarretacasalengua17a}
\begin{align}
    n_e &= \frac{N_e\tau_R}{\eta_e}\\
    P_{0,e} &= 1- P_{1,e} - P_{2,e}\label{eq:p0_exp}\\
    P_{1,e} &= n_e -n^2_e g^{(2)}_{ee}(0) \\
    P_{2,e} & = \frac{n_e^2 g^{(2)}_{ee}(0)}{2}\label{eq:p2_exp}. 
\end{align}

\begin{figure*}
  \centering
  \includegraphics[width=\linewidth]{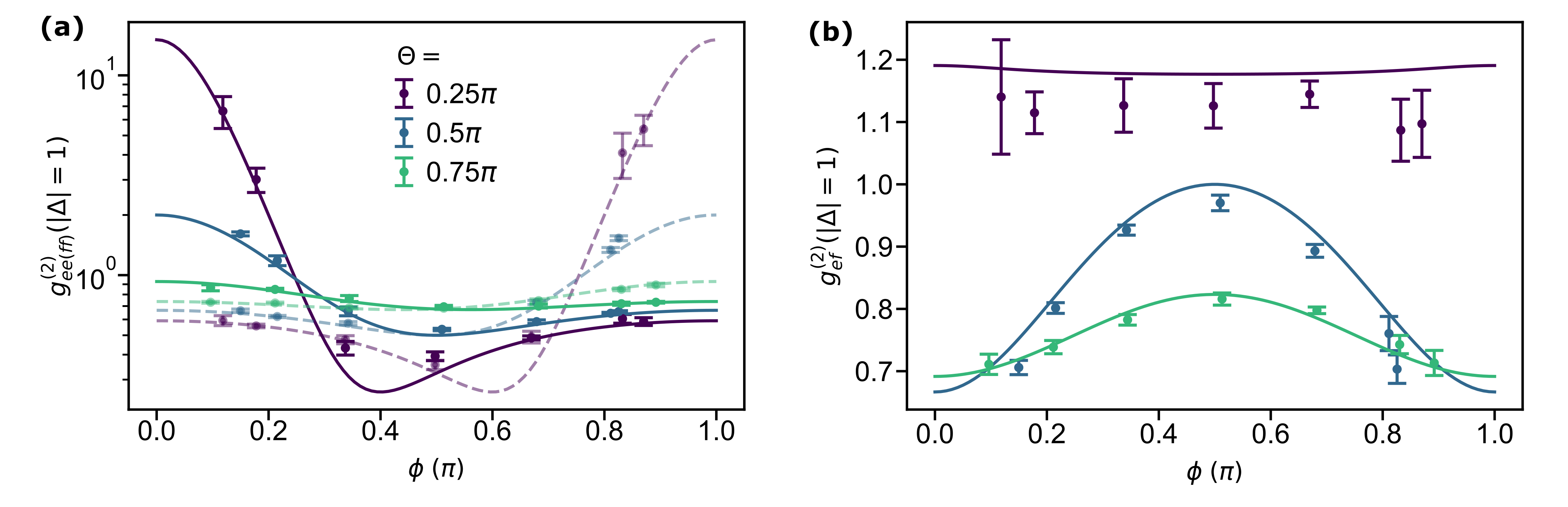}
  \caption{
    Time-integrated correlations at the first neighboring peak ($|\Delta| = 1$).
    (a) Phase-resolved, time-integrated second-order autocorrelations $g^{(2)}_{ee}(|\Delta| = 1)$ (solid symbols, solid curves) and $g^{(2)}_{ff}(|\Delta| = 1)$ (semitransparent symbols, dashed curves).
    (b) Corresponding cross-correlation $g^{(2)}_{ef}(|\Delta| = 1)$. All traces are plotted versus interferometer phase $\phi$ for pulse areas $\Theta = 0.25\pi$ (purple), $0.50\pi$ (blue), and $0.75\pi$ (green).
    Symbols are experimental data; curves are theoretical predictions, confirming an excellent agreement between theory and experiment and further validating the time-bin model.
    }
  \label{fig:tau1}
\end{figure*}

\section{Autocorrelations of MZI output under $\pi$-pulse excitation}
\label{sec:mzi_pi}
Fig.~\ref{fig:MZI_pi_g3_g3} (a) and (b) display the MZI autocorrelations $G^{(2)}_{ee(ff)}(\tau)$ at output $e$($f$) measured under $\pi$-pulse excitation.
At this driving strength, each excitation prepares a nearly pure one‑photon Fock state, eliminating any phase dependence in both the population and the correlation of the MZI outputs.
Accordingly, we conduct the measurement without phase stabilization and observe near‑unity zero‑delay coincidences, yielding $g^{(2)}_{ee}(0)=0.977\pm0.009$ and $g^{(2)}_{ff}(0)=0.984\pm0.009$. 

Given the low multiphoton events of the seed state ($g^{(2)}_{\mathrm{HBT}}(0)<0.006$) shown in Fig.~\ref{fig:QD} (a), the three-photon probability of the MZI outputs is expected to be extremely small. To confirm this, we directly probe the third-order correlation at output $e$ and bound the probability of events with more than two photons. In the configuration used here---resonant excitation of a two-level system with pulses with fixed temporal duration and pulse area $\Theta \in [0,\pi]$---events with $>2$ photons at the output are most probably when the seed state has the maximal multiphoton error for $>1$ photons, i.e., at $\Theta = \pi$~\cite{Fischer2017}. Accordingly, it suffices to examine this particular case to set an upper bound on undesirable multiphoton events with $>2$ photons. For the third-order correlation measurement, we use an extended HBT setup comprising two cascade 50:50 fiber-based beam splitters with three outputs, routing signals to detector 0, 1, and 2. In Fig.~\ref{fig:MZI_pi_g3_g3} (c), we show the unnormalized correlation $G_{e}^{(3)}(\tau_1, \tau_2)$ at output $e$ under $\pi$-pulse excitation. The delays $\tau_{1(2)} = t_{1(2)}-t_0$ represent the time differences between the detection time at detector 1(2) and that at detector 0. To evaluate the correlations between time bins, we integrate the results along both time axes over the pulse repetition period $\tau_R$. The integrated boundaries for each time-bin are indicated by gray lines. As expected for $P_i =0$ of $i\geq3$, no triple coincidences are recorded at the center of the correlation map, $\SI{-6.5}{\ns} \lesssim \tau_{1,2} \lesssim \SI{6.5}{\ns}$, over the entire about $30$-hour acquisition. With a mean of $\approx\!11$ uncorrelated events at $|\tau_{1,2}|\gg0$, we set an upper bound of the normalized, integrated third-order correlation $g_{e}^{(3)}(0)<0.09$, by conservatively assuming one event in the zero-delay bin. 

Given the correlation results above, we estimated the photon-number probabilities at output $e$ for $\Theta=\pi$. The experimental values are obtained using Eqs.~\eqref{eq:p0_exp}--\eqref{eq:p2_exp}, while the theoretical values are calculated with Eqs.~\eqref{eq:p0}--\eqref{eq:p2} for an ideal single-photon emitter. We obtain $P_0 = 0.637$ ($0.625$), $P_1 = 0.254$ ($0.250$), and $P_2 = 0.109$ ($0.125$) for the experimental (theoretical) results. The smaller two-photon probabilities and the larger vacuum and one-photon probabilities are primarily due to the mean photon number being $0.47$ rather than the ideal value $0.50$. As shown above, contributions with more than two photons are negligible. The corresponding upper bound on the three-photon probability $P_3 < 0.0016$, evaluated under the assumption of $P_i =0$ for $i\geq4$~\cite{zubizarretacasalengua17a}.
We note that the dual‑source HOM configuration illustrated in the inset of Fig.~\ref{fig:prob}(d) is expected to yield the same zero-delay correlation results at the same pulse area, but a distinct probability distribution: $P_{0,\mathrm{HOM}}=0.5$, $P_{1,\mathrm{HOM}}=0$, and $P_{2,\mathrm{HOM}}=0.5$ at both outputs.
The identical value $g^{(2)}_{ee(ff)}(0)\approx1$ obtained for these two configurations results from distinct underlying photon-number landscapes, underscoring the non‑monotonic relationship between photon‑number probabilities and second‑order correlations.

\section{Time-Bin Correlations at $|\Delta|=1$}
In the main manuscript, we mainly discuss about correlations at zero-delay, nevertheless, our model reproduces every other time-bin correlation.  
In particular, the $|\Delta| = 1$ correlations in Eqs.~\eqref{eq:auto_1} and \eqref{eq:cross_1} carry both $\phi$- and $2\phi$-dependent contributions.
Fig~\ref{fig:tau1}(a) and (b) display phase-resolved autocorrelations and cross-correlations in the $(\phi,\Theta)$ parameter space.
We compare the theoretical predictions (curves) for the time-bin correlations $g^{(2)}_{ee(ff)}(|\Delta|=1)$ and~$g^{(2)}_{ef}(|\Delta|=1)$ against experimental data (symbols) for pulse areas of $0.25\pi$, $0.50\pi$, and $0.75\pi$. 
We observe a stronger phase impact on the autocorrelations than on the cross-corelation.
At the lowest pulse area (purple), the autocorrelation at output $e$ transits from $g^{(2)}_{ee}(|\Delta|=1)=6.6\pm1.2$ to $0.58\pm0.03$ by varying the phase from $0.12\pi$ to $0.87\pi$. 
while the cross-corelation at $|\Delta|=1$ stays almost unaffected agreeing with the previously reported relation in Ref.~\cite{MaillettedeBuyWenniger24}.
The good agreement between our analytical model and experimental results validates the accuracy of our model across the full range of parameter space.
This theoretical framework not only reproduces observed correlation patterns but also provides insight into the temporal structure of photon statistics such as time-bin entanglement in pulse-train operation. 

\bibliographystyle{apsrev4-2}
\bibliography{bibliography}

\end{document}